\documentclass[12pt]{article}

\usepackage{graphicx}
\usepackage{scicite}

\usepackage{times}

\def\gapp{\ifmmode\stackrel{>}{_{\sim}}\else$\stackrel{>}{_{\sim}}$\fi}
\def\lapp{\ifmmode\stackrel{<}{_{\sim}}\else$\stackrel{>}{_{\sim}}$\fi}

\newenvironment{sciabstract}{%
\begin{quote} \bf}
{\end{quote}}

\newcounter{lastnote}
\newenvironment{scilastnote}{%
\setcounter{lastnote}{\value{enumiv}}%
\addtocounter{lastnote}{+1}%
\begin{list}%
{\arabic{lastnote}.}
{\setlength{\leftmargin}{.22in}}
{\setlength{\labelsep}{.5em}}}
{\end{list}}

\title{Enhanced Optical Emission During Crab Giant Radio Pulses}

\author
{A. Shearer$^{1\ast}$, B. Stappers$^{2,3}$, P. O' Connor$^1$, A. Golden$^1$, \\ 
R. Strom$^{2,3}$, M. Redfern$^1$ \& O. Ryan$^1$ \\
\normalsize{$^1$National University of Ireland, Galway,}\\
\normalsize{ Newcastle Rd., Galway, Ireland}\\
\normalsize{$^{2}$Stichting ASTRON, 7990 AA Dwingeloo, The Netherlands}\\
\normalsize{$^{3}$Sterrenkundig Instituut ``Anton Pannekoek'',}\\
\normalsize{1098 SJ Amsterdam, The Netherlands}\\
\\
\normalsize{$^\ast$To whom correspondence should be addressed; E-mail:  andy.shearer@nuigalway.ie}
}

\date{}

\begin{document} 


\baselineskip24pt


\maketitle


\begin{sciabstract}

We detected a correlation between optical and giant radio pulse emission
from the Crab pulsar. Optical pulses coincident with the giant radio
pulses were on average 3\% brighter than those coincident with normal
radio pulses. Combined with the lack of any other pulse
profile changes, this result indicates that both the giant radio pulses and
the increased optical emission are linked to an increase in the
electron-positron plasma density. 


\end{sciabstract}

Despite more than 30 years of observation the emission mechanism of
pulsars is still a matter of debate\cite{ly99}. A broad consensus does
exist: that luminosity is powered by the rotation of the pulsar, that the
pulsed radio signal comes from a coherent process, and that the optical-to-X-ray 
emission is incoherent synchrotron radiation whereas the
$\gamma$-ray emission is curvature radiation\cite{synch}. What is not
agreed on is the mechanism that accelerates the electrons to the
energy required for synchrotron and curvature radiation, where this
acceleration takes place, how coherency is maintained, and the
stability of the electron-positron plasma outflow from the neutron
star's surface \cite{ha98,ro96}. From the radio-pulse profile at 1380
MHz and the optical profile for the Crab pulsar (Fig. 1), we can identify
two primary features:  a main pulse and an inter-pulse. At lower
energies a radio precursor can be seen and at higher energies in the
optical, x-ray and $\gamma$-ray regions, bridge emission can be seen
between the main pulse and the interpulse. One suggestion is that the
precursor represents emission from the pulsar polar cap region near
the neutron star surface, similar to the radio emission from most
pulsars, and that the other features come from higher in the
magnetosphere\cite{ro95}. On a pulse-by-pulse basis, the radio emission
is chaotic, whereas over a prolonged period the average radio pulse
profile is steady. This picture is made more complex by the existence
of giant radio pulses (GRPs) that occur at random intervals, in phase
with either the main pulse or interpulse (Fig. 1), and that have energies
about 1000 times as high as than the mean energy\cite{lu95}. In the
optical and infrared energy regions the pulse profile is constant at
the 1\% level \cite{heg71,jon80}.

\begin{figure}[h] 
\centering 
\includegraphics[scale=0.9]{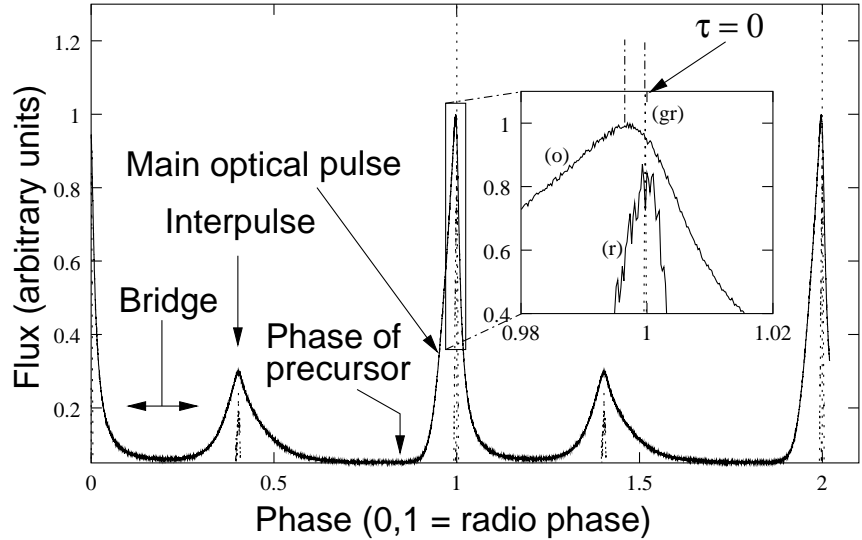}
\caption{The Crab pulse profile showing the optical light curve(o),
the average radio light curve at 1380 MHz(r) and a single giant
pulse at 1357.5MHz (gr). $\tau$, time. Two periods are shown for clarity. Various pulse
parameters have been identified. Also shown is the location of the
precursor observed at lower frequencies and the bridge emission seen
particularly at higher frequencies. On this scale, the GRP
width corresponds to $\approx$ 0.00035 units of phase (12
$\mu$s), the radio pulse to $\approx$ 0.009 (300 $\mu$s), and the optical
pulse to $\approx$ 0.045 (1500 $\mu$s).  The optical light curve is taken from
the second night of observation. The avalanche photodiode (APD)
bandpass for these observations was from 6000 to 7500\AA.  Phase 0
corresponds to the arrival at the solar system barycentre of the peak
radio pulse. The optical light curve for this plot was divided into 5000 phase
bins; the optical peak is at -100 $\mu$s with respect to the JBE.  }
\end{figure}

The Crab pulsar was first detected through its GRP
component\cite{st68}. Since then, only three other pulsars have shown similar
behaviour.  Two of the others are recycled millisecond pulsars -
PSR B1937+21\cite{co96} and PSR B1821-24\cite{ro01} - of which the
latter is in a globular cluster. In comparison with the Crab pulsar
they are much older and have surface magnetic fields some 1/10,000th as strong.
The third, which has been reported only recently, is PSR
B0540-69 \cite{jo03} - a pulsar similar to the Crab pulsar in the
large Magellanic Cloud.  The only feature that seems to link these
pulsars is the magnetic field strength at the light
cylinder\cite{lyc,ro01}. To investigate this possibility and to examine other aspects of
GRP emission, and hence radio emission in general, we would need to
see whether the GRP phenomena can be seen at other energies. Any observed
variation in the emitted flux, pulse morphology, or phase relations at
higher energies coincident with a GRP would provide
explicit constraints on pulsar (coherent/incoherent) emission physics
and geometry. A previous study \cite{lu95} searched for correlations
between low-energy $\gamma$ ray emission and GRPs. Those investigators measured a
$\gamma$-ray giant pulse upper limit of 2.5 times the average flux and
found that no other $\gamma$-ray pulse parameters (width, arrival time,
or spectrum) varied during GRP events.  They estimated that 60\% of
the upper limit could be due to inverse Compton scattering of the
radio photons on the local plasma.

\begin{table}[h]
\begin{center}
\begin{tabular}{c|cc|cc}
\hline 
                  &   \multicolumn{2}{c|} {Start time (UTC)} & \multicolumn{2}{c}{Duration (seconds)} \\ 
Date              &    Optical         & Radio    &    Optical &       Radio  \\ \hline 
2 November 1999  &    23:29:39	& 23:30:00 &	6500.0	&	5400.0 \\
3 November 1999  &    23:30:00	& 23:37:20 &	5320.0	&	6000.0 \\ \hline
\end{tabular}
\end{center}
\caption{Summary of the radio and optical observations}
\end{table}

To investigate whether there is a link between the radio and optical
emission from the Crab pulsar, we made simultaneous observations with
the Westerbork Synthesis Radio Telescope (WSRT) and with the
Transputer Instrument for Fast Image Detection (TRIFFID) 
optical photometer \cite{bu02} mounted on the 4.2 m
William Herschel Telescope (WHT) (Table 1). The radio observations were made
at a central frequency of 1357.5 MHz with the WSRT in tied-array
mode - making it equivalent to a 94-m dish - and the pulsar signal
analyser PuMa (pulsar machine)(\cite{vo02}). A 5 MHz band was Nyquist sampled and
coherently dedispersed\cite{ha71,disp} at a dispersion measure of
56.791 cm$^{-3}$ pc.  Samples were then combined to produce a time
series with a resolution of 6.4 $\mu$s. This time series was searched
for signal-to-noise peaks greater than 10-$\sigma$, equivalent to a
flux density of 150 jansky (Jy) and $\sim1000$ times as strong as the normal
radio emission\cite{back}. All 10,034 peaks were located at pulse longitudes
corresponding to the main pulse and the interpulse, as defined by the average
radio emission, thereby identifying them as GRPs. The energy of the
GRPs was subsequently calculated by summing together all significant
samples within $\pm1.5$\,ms and using the same flux scale given above
for the peaks. The amplitude distribution of the GRPs showed the
expected functional form (Fig. S1) \cite{lu95}. Conditions on the
second night of optical observations were good: Seeing was better
than 1'' for the duration of the observations,  and the transparency
varied by less than 20\%. On the first night, sky transparency was poor
($<$ 50\% of the first night and variable) and the seeing was 1.5''.

Time alignment of the optical and radio data was achieved with the
Jodrell Bank Crab pulsar ephemeris (JBE, \cite{ro02}) appropriate for
November 1999 (Table 2) and the optical and radio arrival times were transformed
to the solar system barycentre with the DE200 Jet Propulsion Laboratory (JPL) ephemeris
\cite{st82}. These data were then folded in phase with the JBE to determine
the relative phase with respect to the radio emission of each optical
photon and each GRP.  After alignment, we found that the average GRP
arrival time is 9.5 $\mu$s before that predicted by the JBE, well
within the 19 $\mu$s error in this ephemeris. The average optical
pulse arrived 100 $\pm$ 20 $\mu$s before the JBE prediction (Fig
1). Those GRPs coincident with the interpulse were coincident with
the peak of the optical interpulse to within the fitting error of
10$\mu$s.

\begin{table}[h]
\begin{center}
\begin{tabular}{llclc} 
\hline
\multicolumn{5}{c}{} \\
\multicolumn{5}{c}{Jodrell Bank Ephemeris valid from MJD 51467.85318 to MJD 51504.74933} \\ \hline    \hline
 Pulse arrival time      &    \multicolumn{4}{l}{$1999/11/02$ 00:00:00.026917   $\pm$   0.000019} UT \\
 Modified Julian Date    &              51484.00000031154      	    & $\pm$   &  0.00000000023 & days \\
 Period                  &           	0.033502974639313  	    & $\pm$   &  0.000000000000192 & s \\
 Pulsar Frequency        &       	29.848095900910629 	    & $\pm$   &  0.000000000170673 & Hz \\
 Frequency 1st derivative  &           -374639.721220770            & $\pm$   &  0.358888933 & $10^{-15} s^{-2}$ \\
 Frequency 2nd derivative  &            10302.350019155             &         &  & $10^{-24} s^{-3}$ \\ [0.8ex] \hline
 \multicolumn{5}{c}{First Pulse Arrival Time (MJD)} \\ \hline \hline
			 &  Optical	        & & 	\multicolumn{2}{c}{Radio} \\
 1999, November 2 :      &  51484.98377619	& &	\multicolumn{2}{c}{51484.98404246} \\
 1999, November 3 :      &  51485.98408505	& &	\multicolumn{2}{c}{51485.98917482} \\ \hline
\end{tabular}
\caption{Ephemeris and pulse arrival times.}
\end{center}
\end{table}

\begin{figure}[h] 
\centering 
\includegraphics[scale=0.95]{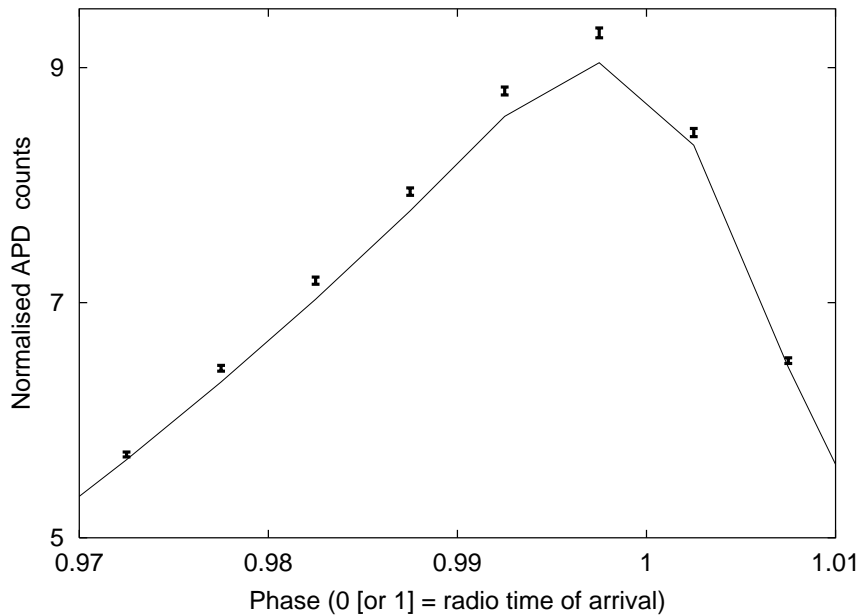}
\caption{The mean optical `giant' pulse superimposed (with its error bars) on
the average optical pulse. The average pulse is determined from the 40
pulsar periods centred on the GRP, but not including it,
from both nights of observation.}
\end{figure} 
 
Because the optical data were subject to fluctuations in sky transparency
(on a timescale of $\approx$ 30 s), only the photons that were
detected 20 periods before and after the GRP and during the GRP
itself, were used for subsequent study. This resulted in a total of
10,034 data sets of 41 periods each. Folding the optical photons at
the Crab's period and then averaging over all data sets (but not
including the period associated with a GRP) we form an average pulse
profile (Fig 2). For comparison we also show the pulse profile formed
by averaging only the optical pulses coincident with a GRP. This
profile shows that the giant optical pulses are on average 3\%
brighter than normal optical pulses (although there is no statistical
difference in the location or sharpness of the peak). We found that
the optical pulses coincident with the GRPs were 7.8 $\sigma$ brighter
than the mean profile (Fig. S2).

We also analyzed other pulse parameters:  arrival time, pulse shape,
and interpulse height. None of these parameters showed any statistically
significant variation with the presence of a GRP (Fig. S3). The
arrival time for the optical giant pulses was on average the same as
the optical pulses coincident with non-GRPs, to within 20 $\mu$s
(Fig 2). The pulse widths for the same pulses were similar at the 10\%
level (Fig. S4). All the individual optical pulse profiles are scaled
down versions of the average optical pulse profile - in contrast to
the radio pulses.

\begin{figure}[h] 
\centering 
\includegraphics[scale=0.95]{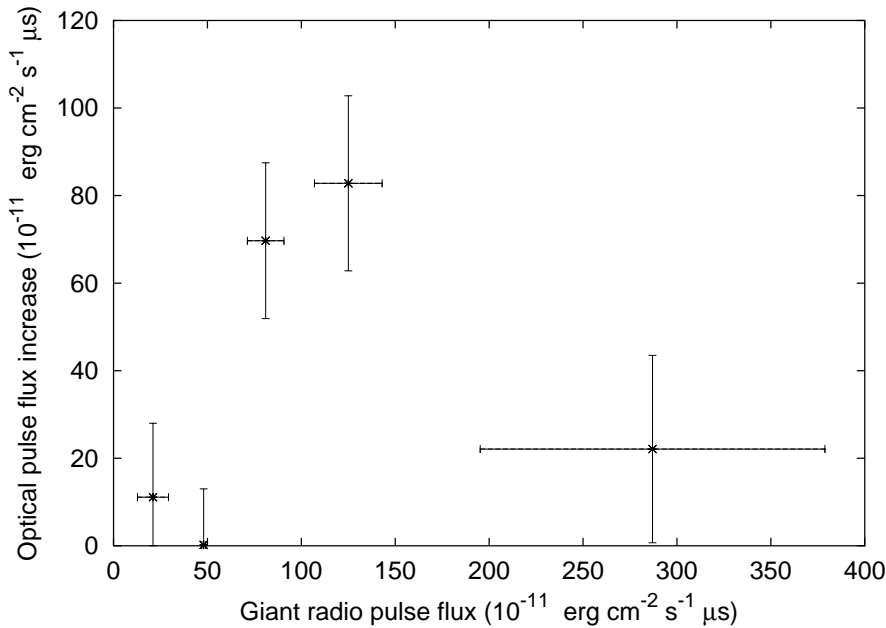}
\caption{The link between optical pulse size and GRP
energy for the second night of observation. The plot shows the
increase in the optical pulse flux against the GRP
flux. The data have been binned using the radio flux, with
approximately equal events in each bin. We calculated the optical pulse flux
assuming that the main peak has an average flux of 1.9 mJy,
\cite{go00} and integrating from 0.9 to 1.05 in phase. The radio flux
was calculated by determining excess counts above noise for $\pm$
0.1 periods around a detected GRP, then converting these counts to janskys.}
\end{figure} 

Some 15\% of the GRPs were coincident in phase with the
interpulse. However, there was no noticeable effect on the optical
interpulse height, giving a 1 $\sigma$ upper limit of 2.5\%. There
was, however, a small effect on the optical main pulse, albeit at low
significance, an increase of 2\% (1.75 $\sigma$). The optical
interpulse showed no change in amplitude during a GRP associated with
the main pulse.  There was a slight increase in the significance of
the main peak variation when only the midrange GRPs were chosen (Fig 3).
To convert the quoted fluxes to luminosity, it would be
necessary to account for possible differences between the radio and
optical beaming fractions\cite{bf}.  From these observations, we have
established a link between optical and radio flux variations.
The fact that only the optical pulse, which is coincident with
a GRP, shows enhanced intensity suggests that the coherent (radio) and
incoherent (optical) emissions produced in the Crab pulsar's
magnetosphere are linked.

Among a number of mechanisms that might explain the radio-optical
connection, several would appear to be unlikely. A decrease in the
opacity affecting the optical luminosity is probably ruled out by the
expected viewing geometry, because the light does not necessarily pass through the
radio emission region. A change in the geometry of the optical
emission region induced by the GRP mechanism is not likely, because we
see no change in optical pulse shape or phase. In particular, we do not
expect the path over which the optical radiation is emitted to have
changed. Increases in magnetic field strength (leading to enhanced
optical emission) can probably be ruled out on a global scale; local
magnetic field enhancements due to increased plasma density might
occur, but they are probably insufficient to explain the
observations. Inverse-Compton scattering of the GRP photons,
kicking them up to the optical band, seems unlikely on several
grounds, because only about 1\% of the radio energy should appear as optical
emission \cite{bl76}, whereas we find nearly equal luminosities. The
$\gamma$-ray upper limit \cite{lu95} also suggests no more than
$\simeq1$\% conversion.

A consistent explanation is that the optical emission is a
reflection of increased plasma density that causes the GRP
event. Whatever triggers the GRP phenomenon, it releases energy
uniformly throughout most of the electromagnetic spectrum, as implied
by the similar energies of radio and enhanced optical pulses.  Changes
in the pair production rate at the level of a few percent could explain the
optical variations and would also be expected at higher energies;
existing limits to enhanced $\gamma$-ray emission \cite{lu95} do not
contradict this. However, an additional mechanism would be needed to
account for the radio GRPs, which are orders of magnitude stronger than
the average pulse level. It has been suggested \cite{bl76} that this
could be achieved by local density enhancements to the plasma stream,
which increase the coherent emission ($\propto n^2$) with little
effect on the (high-energy) incoherent radiation($\propto n$). These
changes must occur on tiny timescales ($<10 \mu$s) to explain
the observed change in optical flux and the upper limit in the
$\gamma$-ray region \cite{lu95,ar79}. This result is also consistent
with the recent observations of nanosecond timescale structure within
GRPs, \cite{ha03}. Whatever the mechanism, our observations
demonstrate a clear link at the individual pulse level between the
coherent and incoherent emission regimes in the Crab pulsar.



\begin{thebibliography}{10}
\bibitem{synch} Synchrotron and curvature radiation are produced when
a charged particle moves in the vicinity of a magnetic
field. Synchrotron radiation is produced when the particle follows a
gyratory path around a magnetic field line and curvature radiation
is the limit when the particle follows a curved field line. See \cite{rl79} for a
detailed treatment.
\bibitem{lyc} The light cylinder is defined as the distance at which
the co-rotation velocity equals the speed of light.
\bibitem{disp} Radio propagation through the interstellar medium is
subject to a frequency-dependent delay that is proportional to the
line-of-sight electron density. This delay can be completely corrected for
 by with the appropriate filter, and in that case it is referred to as coherent dedispersion.
\bibitem{back} The width of the WSRT tied-array fan beam is $\sim10$
arcseconds and the total contribution of the Crab nebula emission to
the system temperature is $\sim120$ K.
\bibitem{bf}
The beaming fraction describes what fraction of the sky is illuminated
by the pulsar during each rotation. It is a reasonable assumption that
this fraction will be energy dependent.
\end{thebibliography}
\bibliographystyle{Science}


\begin{scilastnote}

\item We thank Enterprise Ireland is thanked for its support under the Basic
Grant Research scheme. P.\,O.\,C. is grateful for support under the HEA 
funded CosmoGrid project.  The WHT is operated on the island of La Palma
by the Isaac Newton Group in the Spanish Observatorio del Roque de los
Muchachos of the Instituto de Astrofisica de Canarias. The WSRT is
operated by ASTRON with financial support from the Netherlands
Organisation for Scientific Research (NWO). We thank M. Roberts from Jodrell
Bank for the provision of the radio ephemeris(JBE), R. Butler
for help in the production of this manuscript and A. Boyle
for her help during the optical observations.

\end{scilastnote}

{\bf Supporting Online Material}\\
www.sciencemag.org/cgi/content/full/301/5632/493/DC1\\
Figs. S1 to S4 \\ 
Accepted 18 June 2003

\end{document}